\documentclass[12pt]{amsart}
\usepackage{amsmath,amssymb}
\usepackage{txfonts}
\usepackage{bm}
\usepackage{cite}
\usepackage{graphicx}
\newtheorem{prop}{Proposition}
\begin{document}
\title{Chaotic Boltzmann machines with two elements}
\author{Hideyuki Suzuki}
\address{Department of Mathematical Informatics,
         The University of Tokyo,
         7--3--1 Hongo, Bunkyo, Tokyo 113--8656, Japan.}
\email{hideyuki@mist.i.u-tokyo.ac.jp}
\date{January 28, 2015}

\thanks{This study was supported by
the Aihara Innovative Mathematical Modelling Project,
JSPS through FIRST Program, initiated by CSTP, Japan}

\begin{abstract}
In this brief note, we show that chaotic Boltzmann machines
truly yield samples from the probabilistic distribution
of the corresponding Boltzmann machines
if they are composed of only two elements.
This note is an English translation (with slight modifications)
of the article originally written in Japanese
[H. Suzuki, Seisan Kenkyu 66 (2014), 315--316].
\end{abstract}

\maketitle

\section{Introduction}

Recently, chaotic Boltzmann machines \cite{Suzuki2013cbm}
were proposed as a deterministic implementation of Boltzmann machines.
The apparently stochastic behavior of chaotic Boltzmann machines
is achieved, without any use of random numbers, by chaotic dynamics
that emerges from pseudo-billiard dynamics.
It was shown numerically that the chaotic billiard dynamics of
chaotic Boltzmann machines can be used for generating sample sequences
from the probabilistic distribution of Boltzmann machines,
and it was successfully applied to other spin models
such as the Ising model and the Potts model \cite{Suzuki2013mc}.

Despite these numerical evidences,
there have been no theoretical proof that chaotic Boltzmann machines
yield samples from the probabilistic distribution of the corresponding
Boltzmann machines.

In this brief note,
as a first step of theoretical approach,
we investigate the simplest system.
Namely, we show that chaotic Boltzmann machines
truly yield samples for the corresponding Boltzmann machines
if they are composed of only two elements.
Although our approach cannot be applied to larger chaotic Boltzmann machines
with more than two elements, we expect that it gives some insights
into the dynamics of larger chaotic Boltzmann machines.

Since the proof is not entirely trivial,
it is considered worth making available on arXiv.
This note is an English translation (with slight modifications)
of the article \cite{Suzuki2014} originally written in Japanese.

\section{Chaotic Boltzmann Machines with Two Elements}

Let $S_1$ and $S_2$ be random variables that take values on $\{0,1\}$.
We assume that the probabilistic model $P[S_1, S_2]$ is given in the form of
conditional probabilities $P[S_1 \mid S_2]$ and $P[S_2 \mid S_1]$,
which is common in Boltzmann machines and many other spin models
in statistical physics.

Let us consider a chaotic Boltzmann machine for the probabilistic model.
In addition to the states $s_1$ and $s_2\in \{0,1\}$ of the elements,
we introduce internal states $x_1$ and $x_2\in [0,1]$ of the elements.
Therefore, the state space of the chaotic Boltzmann machine is
$\{0,1\}^2\times[0,1]^2$.
The internal states $x_1$ and $x_2$ evolve
according to the following differential equations:
\begin{align}
\frac{dx_1}{dt} &= (1-2s_1) C_1(s_2) P[s_1\mid s_2]^{-1}, \label{eq:dx1}\\
\frac{dx_2}{dt} &= (1-2s_2) C_2(s_1) P[s_2\mid s_1]^{-1}, \label{eq:dx2}
\end{align}
where $C_i(\cdot)>0$ represents arbitrary positive constants
defined for each state of the other element.
When the internal state $x_i$ reaches $0$ or $1$,
the state $s_i$ of the element changes as follows:
\begin{align}
&s_i \longleftarrow 1 \qquad \text{when} \qquad x_i = 1, \label{eq:ds1}\\
&s_i \longleftarrow 0 \qquad \text{when} \qquad x_i = 0. \label{eq:ds0}
\end{align}
The internal state $(x_1,x_2)$ moves straight inside the unit square $[0,1]^2$,
and changes its direction when it hits a side of the square.
Therefore, this system can be understood as a pseudo-billiard
in the billiard table $[0,1]^2$.

What we expect here is that the state variables $s_1$ and $s_2$ observed from
this system follow the distribution $P[S_1,S_2]$.
Namely, our goal in this note is to show the following proposition.
\begin{prop}
For any initial values
$(s_1(0), s_2(0), x_1(0), x_2(0))\in\{0,1\}^2\times[0,1]^2$,
the chaotic Boltzmann machine defined by Eqs.~(\ref{eq:dx1})---(\ref{eq:ds0})
satisfies
\begin{equation}
\lim_{T\to\infty}\frac{1}{T}\int_0^{T}
  \delta(s_1(t),s'_1)\,\delta(s_2(t),s'_2)\,dt
  = P[s'_1,s'_2]
\label{eq:pss}
\end{equation}
for $(s'_1,s'_2)\in \{0,1\}^2$,
provided that $\displaystyle\frac{R_2(0)+R_2(1)}{R_1(0)+R_1(1)\mathstrut}$ is irrational.
Here, $\delta(\cdot,\cdot)$ is Kronecker's delta function,
and $R_i(s_i)=C_{3-i}(s_i) P[s_i]$.
\end{prop}

\section{Proof: Quasi-Periodic Dynamics}

\begin{figure}
\begin{center}
\includegraphics[scale=0.9]{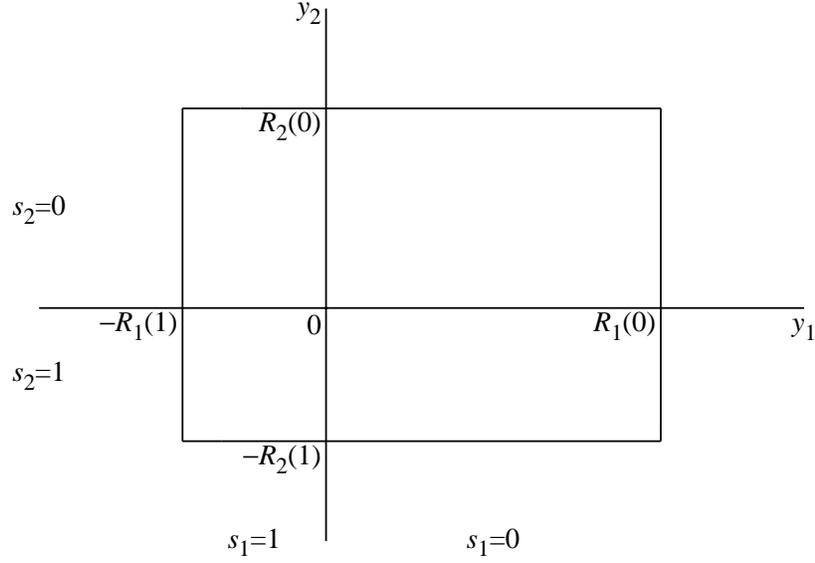}
\caption{\label{fig:sspace}
The state space for $(y_1,y_2)$.}
\end{center}
\end{figure}

To show this proposition, we rewrite
Eqs.~(\ref{eq:dx1}) and (\ref{eq:dx2})
of the chaotic Boltzmann machine.
Specifically, instead of the state $(s_i,x_i)$ of each element,
we introduce a new state variable $y_i=H_i(s_i,x_i)$ using the map
\begin{equation}
H_i(s_i,x_i) = (1-2s_i) R_i(s_i) x_i.
\end{equation}
Then the state space $\{0,1\}^2\times[0,1]^2$ of the chaotic Boltzmann machine
is mapped to the rectangular
$[-R_1(1),R_1(0)]\times[-R_2(1),R_2(0)]$ (see Fig.~\ref{fig:sspace}).
On this state space, the dynamics of the chaotic Boltzmann machine
(Eqs.~(\ref{eq:dx1})--(\ref{eq:ds0}))
is rewritten as follows:
\begin{align}
\frac{dy_1}{dt} &=  R_1(s_1) C_1(s_2) P[s_1|s_2]^{-1}
= \frac{R_1(s_1)R_2(s_2)}{P[s_1,s_2]}, \label{eq:dy1} \\
\frac{dy_2}{dt} &=  R_2(s_2) C_2(s_1) P[s_2|s_1]^{-1}
= \frac{R_1(s_1)R_2(s_2)}{P[s_1,s_2]}, \label{eq:dy2}
\end{align}
and 
\begin{align}
&y_i \longleftarrow -R_1(1) \qquad \text{when} \qquad y_i = R_1(0),\\
&y_i \longleftarrow -R_2(1) \qquad \text{when} \qquad y_i = R_2(0),
\end{align}
where the state $s_i$ can be determined from $y_i$ as
$s_i(y_i)=0$ when $y_i\ge0$, and $s_i(y_i)=1$ when $y_i<0$.

The differential equations (\ref{eq:dy1}) and (\ref{eq:dy2})
always satisfy $\displaystyle\frac{dy_1\mathstrut}{dt\mathstrut}=\frac{dy_2}{dt}$.
Hence, the state $(y_1,y_2)$ moves in the direction of $(1,1)$
in the rectangular state space.
Here we introduce $(z_1,z_2)$ which moves along the orbit of
$(y_1,y_2)$ at a constant velocity
$\displaystyle\frac{dz_1\mathstrut}{dt\mathstrut}=\frac{dz_2}{dt}=1$.
Then the orbit of $(z_1,z_2)$ is equidistributed in the state space
in the following sense.
Let us consider the Poincar\'e section on $z_1=-R_1(1)$.
While the value of $z_1$ changes by $(R_1(0)+R_1(1))$,
the value of $z_2$ changes by the same amount.
Therefore, the Poincar\'e map on $z_1=-R_1(1)$ is a rigid rotation
with rotation number $(R_2(0)+R_2(1))/(R_1(0)+R_1(1)) \bmod 1$.
If the rotation number is irrational, the Poincar\'e map is ergodic
with respect to the uniform distribution.
Then the orbit is also equidistributed in the state space,
and we have
\begin{equation}
\lim_{T\to\infty}\frac{1}{T}\int_0^{T}
  \delta(s_1(z_1(t)),s'_1)\,\delta(s_2(z_2(t)),s'_2)\,dt
  \propto R_1(s'_1)R_2(s'_2).
\end{equation}
Since $(y_1,y_2)$ moves along the same orbit as $(z_1,z_2)$
with the velocity described in Eqs.~(\ref{eq:dy1}) and (\ref{eq:dy2}),
we have
\begin{equation}
\lim_{T\to\infty}\frac{1}{T}\int_0^{T}
  \delta(s_1(y_1(t)),s'_1)\,\delta(s_2(y_2(t)),s'_2)\,dt
  = P[s'_1,s'_2],
\end{equation}
which completes the proof.

\section{Concluding Remarks}

In this brief note, we have shown that chaotic Boltzmann machines
with two elements have quasi-periodic dynamics and
yield samples from the probabilistic distribution
of the corresponding Boltzmann machines,
provided that the rotation number is irrational.

The set of parameter values $C_i(\cdot)$
that make rotation numbers rational has Lebesgue measure zero.
This is analogous to probabilistic Monte Carlo sampling,
which does not work with probability (measure) zero.
However, when we design chaotic Boltzmann machines artificially,
it may be possible that the rotation number easily becomes rational.
In such a case, we have to adjust $C_i(\cdot)$ by multiplying some constants
to make the rotation number irrational.

The performance as an sampling algorithm
depends on the characteristics of the rotation number.
At least, a chaotic Boltzmann machine with a good rotation number
that yields a low-discrepancy sequence is expected to exhibit
better performance than the corresponding Boltzmann machine.

Our approach in this note cannot be applied
to larger chaotic Boltzmann machines with more than two elements,
which exhibit chaotic behavior.
However, some aspects are expected to be shared with larger systems;
for example, our preliminary numerical study shows that irrationality is
important, for chaotic Boltzmann machines
composed of not many but more than two elements,
to generate faithful sample sequences.
Therefore, we expect that our approach gives some insights
into the dynamics of larger chaotic Boltzmann machines.

\end{document}